\documentclass[slac_one]{revtex4}
\usepackage{graphicx}
\usepackage{fancyhdr}
\renewcommand{\headrulewidth}{0pt}
\renewcommand{\footrulewidth}{0pt}

\def\beq   {\begin{equation}}
\def\eeq   {\end{equation}}
\def\beqd  {\begin{displaymath}}
\def\eeqd  {\end{displaymath}}
\def\beqaa {\begin{eqnarray}}
\def\eeqaa {\end{eqnarray}}

\def\ti  {\tilde}

\def\sf  {\ti f}

\def\sn  {\ti \nu}
\def\sl  {\ti \ell}

\def\nt  {\tilde\chi^0}
\def\ch  {\tilde\chi^\pm}
\def\chp {\tilde\chi^+}

\def\a   {\alpha}
\def\b   {\beta}
\def\t   {\theta}

\def\sz{\ifmmode{\tilde{\chi}^0} \else{$\tilde{\chi}^0$} \fi}
\def\sw{\ifmmode{\tilde{\chi}} \else{$\tilde{\chi}$} \fi}

\begin{document}

\title{Impact of slepton generation mixing on the search for sneutrinos} 

%

\author{K. Hidaka}
\affiliation{Department of Physics, Tokyo Gakugei University, Koganei,
Tokyo 184-8501, Japan}
%

\begin{abstract}
We perform a systematic study of sneutrino ($\sn$) production and decays in the 
Minimal Supersymmetric Standard Model (MSSM) with lepton flavour violation (LFV).
We study bosonic decays of $\sn$ as well as fermionic ones. We show that the effect 
of slepton generation mixing on the $\sn$ production and decays can be quite large 
in a significant part of the MSSM parameter space despite the very strong experimental 
limits on LFV processes. This could have an important impact on the search for 
$\sn$'s and the determination of the MSSM parameters at LHC and future colliders, 
such as ILC, CLIC and muon collider.
\end{abstract}

\maketitle

\thispagestyle{fancy}


\section{INTRODUCTION \label{intro}} 
Systematic studies of decays of sneutrinos, the supersymmetric (SUSY) 
partners of neutrinos, in the Minimal Supersymmetric Standard Model 
(MSSM) have been performed already \cite{Bartl: slepton decay}.
In these studies it is assumed that there is no generation mixing in 
the slepton sector. 
In this article based on \cite{Bartl: LFVsnu} we study the effect of 
slepton generation mixing on the production and decays of the sneutrinos 
in the MSSM. 
Lepton flavour violating (LFV) productions and decays of SUSY particles 
have been studied for the case of slepton generation mixing \cite{LFV_refs}. 
Some of the studies are rather model dependent. Furthermore, so far no 
systematic study of LFV in sneutrino decays including bosonic decays 
has been performed. The aim of this article is to perform a systematic study 
of sneutrino production and decays including the bosonic decay modes in 
the general MSSM with LFV in slepton sector. 

\section{THE MODEL \label{model}}
  First we summarize the MSSM parameters in our analysis. 
The most general charged slepton mass matrix including left-right mixing
as well as flavour mixing in the basis of
$\sl_{0\a}=(\tilde e_L,\tilde\mu_L,\tilde\tau_L,\tilde
e_R,\tilde\mu_R,\tilde\tau_R)$, $\a=1,...,6$,  
is given by \cite{Bartl: LFVsnu}:
%
\[
M^2_{\tilde \ell} = \left(
\begin{array}{cc}
M^2_{LL} &  M^{2\dagger}_{RL} \\
M^2_{RL} &  M^2_{RR} \\
\end{array} \right)~,
\]
%
with
\begin{eqnarray*}
M^2_{LL,\alpha\beta} &=& 
M^2_{L,\alpha\beta} + m^2_Z\cos(2\b)(-\frac{1}{2}+\sin^2{\t_W})\delta_{\a\b}
                    + m^2_{\ell_\a}\delta_{\a\b} \, ,\\
%
M^2_{RR,\a\b} &=& M^2_{E,\a\b}-m^2_Z\cos(2\b)\sin^2{\t_W}\delta_{\a\b} 
+m^2_{\ell_\a}\delta_{\a\b} \, ,\\
%
M^2_{RL,\a\b} &=& v_1 A_{\b\a}-m_{\ell_\a}\mu^*\tan\b\delta_{\a\b} \, .
\end{eqnarray*}
The indices $\a,\b=1,2,3$ characterize the flavours $e,\mu,\tau$, respectively.
$M^2_{L}$ and $M^2_{E}$ are the hermitean soft SUSY breaking mass matrices for
left and right sleptons, respectively. 
$A_{\a\b}$ are the trilinear couplings of the sleptons and the Higgs boson.
$\mu$ is the higgsino mass parameter.
$v_1$ and $v_2$ are the vacuum expectation values of the Higgs fields
with $v_1=\langle H^0_1\rangle$, $v_2=\langle H^0_2\rangle$, 
and $\tan\b\equiv v_2/v_1$. 
$m_{\ell_\a}$ are the physical lepton masses. 
The physical mass eigenstates $\sl_i$, $i=1,...,6$, are given by 
$\sl_i = R^{\sl}_{i\a} \sl_{0\a}$. 
The mixing matrix $R^{\sl}$ and the physical mass eigenvalues
are obtained by an unitary transformation 
$R^{\sl}M^2_{\sl}R^{\sl\dagger}=
{\rm diag}(m^2_{\sl_1},\dots,m^2_{\sl_6})$, 
where $m_{\sl_i} < m_{\sl_j}$ for $i<j$.
Similarly, the mass matrix for the sneutrinos, in the basis
$\sn_{0\a}=(\tilde\nu_{eL},\tilde\nu_{\mu L},\tilde\nu_{\tau L})\equiv
(\tilde\nu_{e},\tilde\nu_{\mu},\tilde\nu_{\tau})$, reads
\begin{eqnarray*}
M^2_{\sn,\a\b} &=&  M^2_{L,\a\b} 
+ \frac{1}{2} m^2_Z\cos(2\b)\delta_{\a\b} 
\quad (\alpha,\beta=1,2,3)~,
\end{eqnarray*}
where the physical mass eigenstates are given by
$\sn_i = R^{\sn}_{i\a}\sn_{0\a}$, $i=1,2,3$,
($m_{\sn_1} < m_{\sn_2} < m_{\sn_3}$).\\
The properties of the charginos $\ch_i$ ($i=1,2$, $m_{\ch_1}<m_{\ch_2}$) 
and neutralinos $\nt_k$ ($k=1,...,4$, $m_{\nt_1}< ...< m_{\nt_4}$)  
are determined by the parameters $M_2$, $M_1$, $\mu$ and $\tan\b$, 
where $M_2$ and $M_1$ are the SU(2) and U(1) gaugino masses, respectively. 
Assuming gaugino mass unification we take $M_1=(5/3)\tan^2\t_W M_2$. 

\section{CONSTRAINTS \label{constraints}}
In our analysis, we impose the following conditions on the MSSM parameter space 
in order to respect experimental and theoretical constraints which are described 
in detail in \cite{Bartl: LFVsnu}:

\renewcommand{\labelenumi}{(\roman{enumi})} 
\begin{enumerate}
  \item The vacuum stability conditions \cite{LFV_refs}, such as \\
        $|A_{\a\b}|^2 < Y_{E,\gamma\gamma}^2(M^2_{L,\a\a} + M^2_{E,\b\b} + m^2_1)$,
        ($\a\neq\b$; $\gamma=Max(\a,\b)$; $\a,\b=1,2,3=e,\mu,\tau$). 
  \item Experimental limits on the LFV lepton decays \cite{LFV_refs}, such as \\
        $B(\mu^- \to e^- \gamma) < 1.2 \times 10^{-11}$,
        $B(\tau^- \to \mu^- \gamma) < 4.5 \times 10^{-8}$, and
        $B(\tau^- \to e^- \gamma) < 1.1 \times 10^{-7}$.
  \item Experimental limits on SUSY contributions to \\
        anomalous magnetic moments of leptons \cite{LFV_refs}
        , e.g.
        $|\Delta a_\mu^{SUSY} - 287 \times 10^{-11}|<178 \times 
        10^{-11}$ (95\% CL).
  \item The LEP limits on SUSY particle masses \cite{LFV_refs}. 
  \item The limit on $m_{H^+}$ and $\tan\b$ from the experimental data on 
        $B(B_u^- \to \tau^- {\bar\nu}_\tau)$ \cite{LFV_refs}.

\end{enumerate}
It has been shown that in general the limit on the
$\mu^-- e^-$ conversion rate is respected if the
limit on $\mu\to e~ \gamma$ is fulfilled \cite{LFV_refs}.
%
The conditions (i)-(v) strongly constrain the MSSM parameters as shown in \cite{Bartl: LFVsnu}.

\section{NUMERICAL RESULTS \label{numerics}}
The possible fermionic and bosonic two-body decay modes of sneutrinos are 
%
%
$\sn_i \rightarrow \nu \nt_j,~ \ell^-_\a \chp_k$ and
$\sn_i \rightarrow \sl^-_j W^+,~ \sl^-_j H^+$, respectively.

We take $\tan\b, m_{H^+}, M_2, \mu, M^2_{L,\a\b}, M^2_{E,\a\b}$, and $A_{\a\b}$ 
as the basic MSSM parameters at the weak scale. We assume them to be real. 
The LFV parameters are $M^2_{L,\a\b}$, $M^2_{E,\a\b}$, and $A_{\a\b}$ with $\a \neq \b$.
We take the following $\ti\mu-\ti\tau$ mixing scenario as a reference scenario with
LFV within reach of LHC and ILC: \\
$\tan\b=20$, $m_{H^+}=150GeV$, $M_2=650GeV$, $\mu=150GeV$, 
$M^2_{L,11}=(430GeV)^2$, $M^2_{L,22}=(410GeV)^2$, $M^2_{L,33}=(400GeV)^2$, 
$M^2_{L,12}=M^2_{L,13}=(1GeV)^2$, $M^2_{L,23}=(61.2GeV)^2$, 
$M^2_{E,11}=(230GeV)^2$, $M^2_{E,22}=(210GeV)^2$, $M^2_{E,33}=(200GeV)^2$, 
$M^2_{E,12}=M^2_{E,13}=(1GeV)^2$, $M^2_{E,23}=(22.4GeV)^2$,
$A_{23}=25GeV, A_{33}=150GeV$, and all the other $A_{\a\b}=0$.\\
In this scenario satisfying all the conditions (i)-(v) above we have:\\
$
m_{\sn_1}=393GeV, m_{\sn_2}=407GeV, m_{\sn_3}=425GeV, \\
\vspace{-0.3cm}\\
\sn_1=-0.36\sn_{\mu}+0.93\sn_{\tau} \sim \sn_{\tau},\\
\sn_2=0.93\sn_{\mu}+0.36\sn_{\tau} \sim \sn_{\mu},\\
\sn_3 \simeq \sn_{e}, \\
\vspace{-0.3cm}\\
m_{\sl_1}=204GeV, m_{\sl_2}=215GeV, m_{\sl_3}=234GeV, \\
\vspace{-0.3cm}\\
\sl_1= -0.0029\ti{\mu}_L+0.033\ti{\tau}_L
-0.12\ti{\mu}_R+0.99\ti{\tau}_R \sim \ti{\tau}_R,\\
\sl_2=0.0022\ti{\mu}_L+0.0040\ti{\tau}_L
+0.99\ti{\mu}_R+0.12\ti{\tau}_R \sim \ti{\mu}_R,\\
\sl_3 \simeq \ti{e}_R,\\
\vspace{-0.3cm}\\
B(\sn_1 \to \mu^- + \chp_1)=0.014, B(\sn_1 \to \tau^- + \chp_1)=0.36, B(\sn_1 \to \sl^-_1 + H^+)=0.48, \\
\vspace{-0.3cm}\\
B(\sn_2 \to \mu^- + \chp_1)=0.20, B(\sn_2 \to \tau^- + \chp_1)=0.12, B(\sn_2 \to \sl^-_1 + H^+)=0.38.
$
\\
\vspace{-0.3cm}\\
As $\sn_2 \sim \sn_\mu$ and $\sl^-_1 \sim \ti\tau^-_R$, the decays 
$\sn_2 \to \tau^- \chp_1$ and $\sn_2 \to \sl^-_1 H^+$ 
are essentially LFV decays. Note that the branching ratios of these LFV decays 
are sizable in this scenario. The reason is as follows: 
The lighter neutralinos $\nt_{1,2}$ and the lighter chargino $\ch_1$ 
are dominantly higgsinos as $M_{1,2} \gg |\mu|$ in this scenario. 
Hence the fermionic decays into 
$\nt_{1,2}$ and $\chp_1$ are suppressed by the small lepton Yukawa 
couplings except for the decay into $\tau^- \chp_1$ which does not 
receive such a suppression because of the sizable $\tau$ Yukawa coupling 
$Y_{E,33}$ for large $\tan\b$. This leads to an enhancement of 
the bosonic decays into the Higgs boson $H^+$. 
Moreover the decay $\sn_2(\sim \sn_\mu) \to \sl^-_1(\sim \ti\tau^-_R)+ H^+$ 
is enhanced by the sizable trilinear $\sn_\mu-\ti\tau^+_R-H^-_1$ coupling 
$A_{23}$ (with $H^-_1=H^-\sin\b$). 
Because of the sizable $\tilde\nu_\mu-\tilde\nu_\tau$ mixing term $M^2_{L,23}$
the $\sn_2$ has a significant $\sn_\tau$ component, which 
results in a further enhancement of this decay due to the large trilinear 
$\sn_\tau-\ti\tau^+_R-H^-_1$ coupling $A_{33}$ ($=150$~GeV). \\
The decays of $\sn_1$ and $\sn_2$ into $\sl^-_{1,2} W^+$ are suppressed since 
$\sl^-_1 \sim \ti\tau^-_R$ and $\sl^-_2 \sim \ti\mu^-_R$.


\subsection{$\tilde\nu$ decay branching ratios \label{BRs}}
We study the basic MSSM parameter dependences of the LFV 
sneutrino decay branching ratios for the reference scenario specified above.
In the following we use the quantities 
$R_{L23}\equiv M^2_{L,23}/$ $((M^2_{L,11}+M^2_{L,22}+M^2_{L,33})/3)$ and 
$R_{A23} \equiv A_{23}/((|A_{11}|+|A_{22}|+|A_{33}|)/3)$ as a measure of LFV.
%
%
In Fig.1 we show the contours of the decay branching ratios 
$B(\sn_2 \to \tau^-  \chp_1)$ and $B(\sn_2 \to \sl^-_1  H^+)$ 
in the $R_{L23}-R_{A23}$ plane, where all basic parameters other than 
$M^2_{L,23}$ and $A_{23}$ are fixed as in the reference scenario specified above. 
As can be seen, these LFV decay branching ratios can be large in a sizable region 
of the $R_{L23}-R_{A23}$ plane and their dependences on $R_{L23}$ and $R_{A23}$ 
are quite remarkable and very different from each other. Hence, a simultaneous 
measurement of these two branching ratios could play an important role in 
determination of the LFV parameters $M^2_{L,23}$ and $A_{23}$. 
Here note that $\sl^-_1 \sim \ti\tau^-_R$ and that the $\sn_\tau$ component in 
$\sn_2(\sim \sn_\mu)$ increases with the increase of the $\ti\nu_\mu - \ti\nu_\tau$ 
mixing parameter $M^2_{L,23}$, which explains the $R_{L23}$ dependence of these 
branching ratios. The decay $\sn_2 \to \sl^-_1  H^+$ can be enhanced also by 
a sizable $A_{23}$ as explained above, which explains the $R_{A23}$ dependence of
$B(\sn_2 \to \sl^-_1  H^+)$. \\
%
\begin{figure}
\includegraphics[width=0.45\textwidth,height=0.28\textwidth,angle=0]{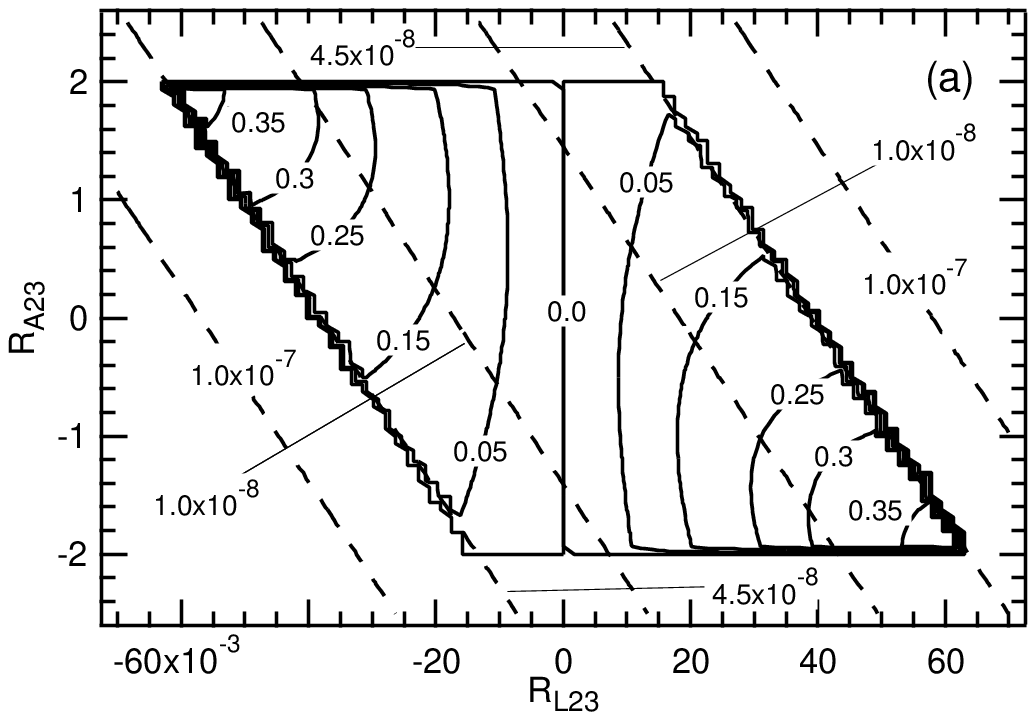}
\includegraphics[width=0.45\textwidth,height=0.28\textwidth,angle=0]{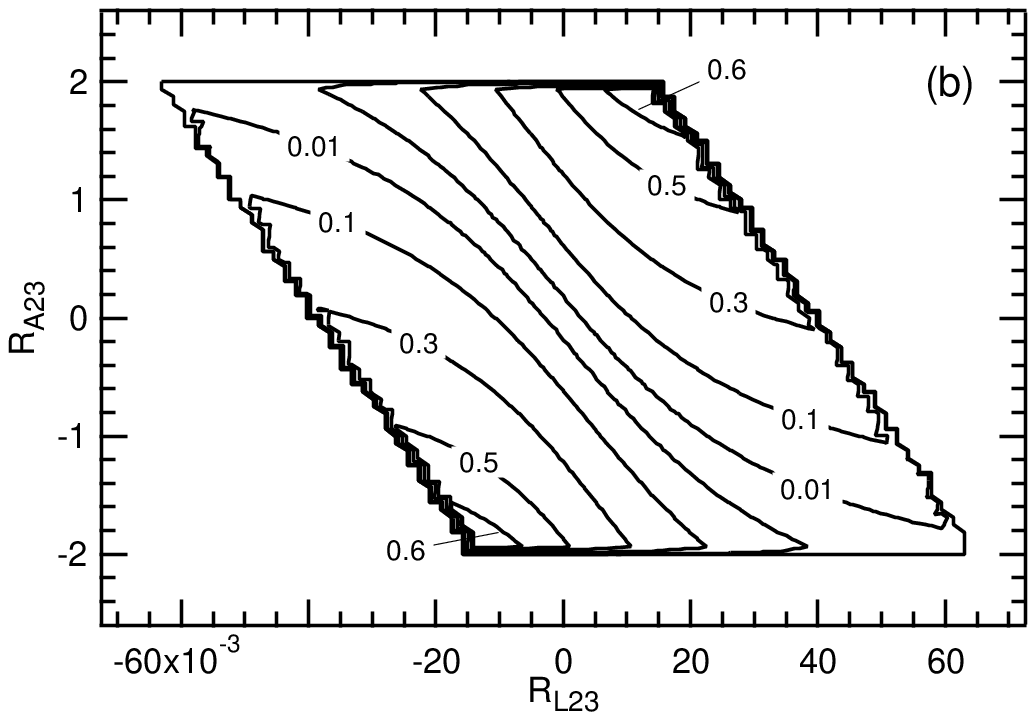}
\caption{
Contours of the LFV decay branching ratios (a) $B(\sn_2 \to \tau^-  \chp_1)$ 
and (b) $B(\sn_2 \to \sl^-_1  H^+)$ in the $R_{L23}-R_{A23}$ plane for our 
$\ti\mu-\ti\tau$ mixing scenario. The region with no solid contours is excluded 
by the conditions (i) to (v) given in the text. The dashed lines in (a) show 
contours of $B(\tau^- \to \mu^- \gamma)$. 
}
\label{fig1}       
\end{figure}
In Fig.2 we show a scatter plot of the LFV decay branching ratios 
$B(\sn_2 \to \sl^-_1  H^+)$ versus $B(\tau^- \to \mu^- \gamma)$ for our 
$\ti\mu-\ti\tau$ mixing scenario with the parameters 
$M_2, \mu, R_{L23}, R_{E23}, R_{A23}$ and $R_{A32}$ varied in the ranges 
satisfying the conditions (i) to (v) given above.
Here $R_{E23}\equiv M^2_{E,23}/$ $((M^2_{E,11}+M^2_{E,22}+M^2_{E,33})/3)$ and 
$R_{A32} \equiv A_{32}/((|A_{11}|+|A_{22}|+|A_{33}|)/3)$. 
All parameters other than $ M_2, \mu, M^2_{L,23}, M^2_{E,23}, A_{23} $ and $A_{32}$ 
are fixed as in the reference scenario specified above. As can be seen in Fig.2, the LFV 
branching ratio $B(\sn_2 \to \sl^-_1  H^+)$ could go up to 30\%
even if the present bound on $B(\tau^- \to \mu^- \gamma)$ improves
by one order of magnitude.
%
\begin{figure}
\includegraphics[width=0.40\textwidth,height=0.25\textwidth,angle=0]{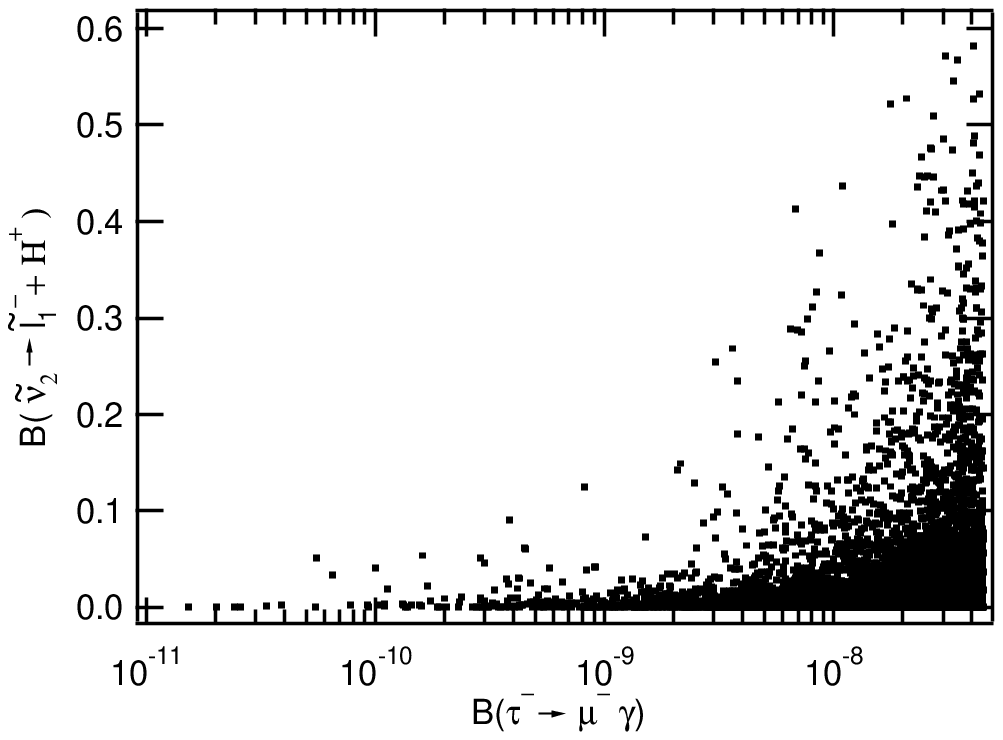}
\caption{
Scatter plot of the LFV decay branching ratios 
$B(\sn_2 \to \sl^-_1  H^+)$ versus $B(\tau^- \to \mu^- \gamma)$ for our 
$\ti\mu-\ti\tau$ mixing scenario. 
}
\label{fig2}       
\end{figure}
For the other LFV decay branching ratios of $\sn_{1,2}$ versus 
$B(\tau^- \to \mu^-  \gamma)$ we have obtained scatter plots similar to Fig.2 
\cite{Bartl: LFVsnu}.
Note here that $B(\sn_1 \to \sl^-_2  H^+)$ can be very large 
due to sizable $M^2_{E,23}$, $A_{32}$ and large $A_{33}$.

We have also studied sneutrino decay branching ratios in the case of 
$\ti e-\tilde\tau$ mixing, where we have obtained similar results to those 
in the case of $\tilde\mu-\tilde\tau$ mixing \cite{Bartl: LFVsnu}. 

\subsection{LFV contributions to collider signatures \label{signatures}}
It is to be noted that in $\ti e-\ti\tau$ mixing scenario the 
t-channel chargino exchanges contribute significantly to the cross 
sections $\sigma(e^+e^-\to\sn_i\bar{\sn}_j)\equiv \sigma_{ij}$ for 
$i,j=$1,3, enhancing the cross sections (including the LFV production 
cross section $\sigma_{13}$) strongly, where $\sn_1 \sim \sn_\tau$ 
and $\sn_3 \sim \sn_e$ \cite{Bartl: LFVsnu}.
We have studied the LFV contributions to signatures of sneutrino
production and decay at the ILC \cite{Bartl: LFVsnu}.
We have shown that the LFV processes (including the LFV $\sn_i$ 
productions and the LFV bosonic $\sn_i$ decays also) can contribute 
significantly to signal event rates. 
%
%
This strongly suggests that one should take into account the 
possibility of the significant contributions of both the LFV 
fermionic and bosonic decays in the sneutrino search and 
should also include the LFV parameters in 
the determination of the basic SUSY parameters at colliders.

\section{SUMMARY \label{summary}}
We have performed a systematic study of sneutrino production and 
decays including both fermionic and bosonic decays in the general MSSM 
with slepton generation mixings. We have shown that LFV sneutrino production 
cross sections and LFV sneutrino decay branching ratios can be quite large 
due to slepton generation mixing in a significant part of the MSSM 
parameter space despite the very strong experimental limits on LFV processes.
This could have an important impact on the search for sneutrinos and the 
MSSM parameter determination at LHC and future colliders, such as ILC, CLIC 
and muon collider.

\begin{acknowledgments}
I deeply thank the other authors of \cite{Bartl: LFVsnu}: A.~Bartl, 
K.~Hohenwarter-Sodek, T.~Kernreiter, W.~Majerotto and W.~Porod. 
\end{acknowledgments}

\end{document}